\documentclass[12pt]{article}	  
\usepackage{graphicx}
\begin{document}
\begin{center}
{\bf Frequency comparisons and absolute frequency measurements of $^{171}$Yb$^+$ single-ion optical frequency standards}\\

\bigskip

E. Peik$^{1,*}$, B. Lipphardt$^1$, H. Schnatz$^1$, T. Schneider$^1$, Chr. Tamm$^1$,\\
S. G. Karshenboim$^{2,3}$\\

\bigskip

$^1$Physikalisch-Technische Bundesanstalt,\\
Bundesallee 100, 38116 Braunschweig, Germany\\
$^2$D. I. Mendeleev Institute for Metrology (VNIIM), 198005 St. Petersburg, Russia\\
$^3$Max-Planck-Institut f\"ur Quantenoptik, 85748 Garching, Germany\\

\bigskip

$^*$ e-mail: ekkehard.peik@ptb.de\\

\end{center}

{\bf Abstract}

We describe experiments with an optical frequency standard based on a laser cooled $^{171}$Yb$^+$ ion confined in a radiofrequency Paul trap. The electric-quadrupole transition from the $^2S_{1/2}(F=0)$ ground state to the $^2D_{3/2}(F=2)$ state at the wavelength of 436 nm is used as the reference transition. 
In order to compare two $^{171}$Yb$^+$ standards, separate frequency servo systems are employed to stabilize 
two probe laser frequencies to the reference transition line centers of two independently stored ions. The  experimental results indicate a relative instability (Allan standard deviation) of the optical frequency difference between the two systems of $\sigma_y(1000\,{\rm s})=5\cdot 10^{-16}$ only, so that shifts in the sub-hertz range can be resolved.  Shifts of several hertz are observed if a stationary electric field gradient is superimposed on the radiofrequency trap field.
The absolute optical transition frequency  of Yb$^+$ at 688 THz was measured with a cesium atomic clock at two times separated by 2.8 years. A temporal variation of this frequency can be excluded within a 
$1\sigma$ relative uncertainty of 
$4.4\cdot 10^{-15}$~yr$^{-1}$. Combined with recently published values 
for the constancy of other transition frequencies
this measurement provides a limit on the present variability of the fine structure constant $\alpha$ at the level of $2.0\cdot 10^{-15}$~yr$^{-1}$.

\newpage

\section{Introduction}

Time and frequency are presently by far the most precisely measurable physical quantities 
and optical atomic frequency standards hold great potential for further improvements in stability and accuracy.
A single trapped and laser cooled ion can be used as a reference that offers excellent control over possible systematic effects.
Presently, a number of groups is pursuing research along these lines with different ions (see \cite{Madej,Gill} for recent reviews). High-resolution spectroscopy and  precise frequency measurements using femto\-second laser frequency comb generators
\cite{Reichert} 
have been performed with reference transitions in Sr$^+$ \cite{Bernard,Margolis}, In$^+$ \cite{Zanthier,Becker}, Yb$^+$ \cite{Stenger1,Tamm1,Tamm2,Blythe} and
Hg$^+$ \cite{Rafac,Bize}. 

In this paper we review two recent experiments at PTB with an optical frequency standard based on the
$^{171}$Yb$^+$ ion: (i)
frequency comparisons between two ions in separate traps are used to investigate the stability and to evaluate
systematic frequency shifts \cite{icols}, (ii) a sequence of absolute optical frequency measurements has been made using a cesium fountain clock in order to test the reproducibility of the standard and to obtain a 
stringent limit on a possible temporal variation of the fine structure constant \cite{peik}.

\section{$^{171}$Yb$^+$ optical frequency standard}

The $^{171}$Yb$^+$ ion is attractive for an optical frequency standard because reference transitions with vanishing low-field linear Zeeman frequency shift are 
available in a level system with relatively simple hyperfine and magnetic sublevel structure (cf. Fig. 1a).
The electric quadrupole transition $(^2S_{1/2},F=0) \rightarrow (^2D_{3/2},F=2)$
at 436 nm wavelength can serve as the reference transition, with a natural linewidth of 3.1 Hz. 
A single ytterbium ion is trapped in a miniature Paul trap and is laser-cooled to a sub-millikelvin temperature
by exciting the low-frequency wing of the quasi-cyclic $(F=1) \rightarrow (F=0)$ component of the
$^2S_{1/2} \rightarrow {^2P_{1/2}}$ resonance transition at 370 nm. A static magnetic field of approximately
$0.3$~mT is applied in order to prevent optical pumping to a nonabsorbing superposition of the magnetic sublevels of the $F=1$ ground state. The natural linewidth of this transition is 21 MHz, which implies a one-dimensional kinetic temperature of 0.6 mK at the Doppler cooling limit. A weak sideband of the cooling radiation provides hyperfine repumping from the $F=0$ ground state to the $^2P_{1/2}(F=1)$ level. At the end of each cooling phase, the hyperfine repumping is switched off in order to prepare the ion in the $F=0$ ground state.
The rapid spontaneous decay from the $^2P_{1/2}$ state to the $(F=2)$ sublevel of the metastable $^2D_{3/2}$ state that occurs during laser cooling is compensated for by coupling this level with an additional laser at 935 nm wavelength to the $[3/2]_{1/2} (F=1)$ state, from where the ion readily returns to the ground state.
The $(F=2)$ sublevel of the $^2D_{3/2}$ state is not rapidly populated or depleted by the laser cooling excitation. Individual quantum jumps to this state due to excitation of the reference transition can therefore be detected using the electron shelving scheme.

\begin{figure}[tbh]
\begin{center}
\includegraphics[width=10cm]{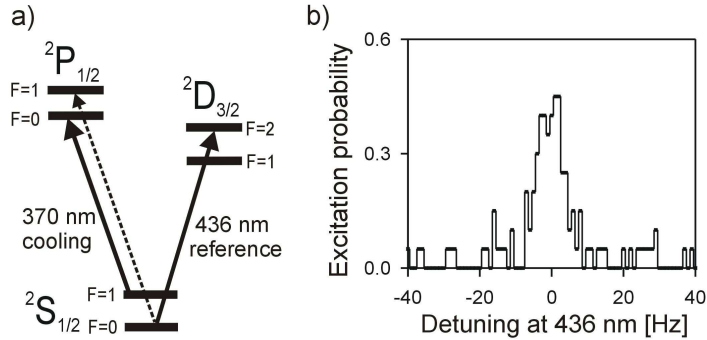}
\end{center}
\caption{a) Simplified level scheme of the $^{171}$Yb$^+$ ion. 
b) Excitation spectrum of the $^2S_{1/2}(F=0,m_F=0)\rightarrow {^2D}_{3/2}(F=2,m_F=0)$ transition,
obtained with 20 probe cycles for each value of the detuning. The linewidth of 10 Hz is approximately at the Fourier limit for the employed probe pulses of 90 ms duration.}
\end{figure}

The $S-D$ electric quadrupole reference transition is probed by the frequency doubled radiation from a diode laser
emitting at 871 nm. The short term frequency stability of this laser is derived from a temperature-stabilized and seismically isolated high-finesse reference cavity.
Figure 1b shows a high-resolution excitation spectrum obtained with 90 ms long laser pulses, leading to an approximately Fourier-limited linewidth in the range of 10 Hz, or a resolution $\Delta \nu/\nu$ of $1.4\cdot 10^{-14}$. Since the duration of the probe pulse is longer than the lifetime of the excited state (51 ms), the observed maximum excitation probability is limited by spontaneous decay.
In order to operate the system as a frequency standard, both wings of the resonance are probed alternately, and the probe light frequency is stabilized to the line center according to the difference of the measured excitation probabilities.

\begin{figure}[tbh]
\begin{center}
\includegraphics[width=9cm]{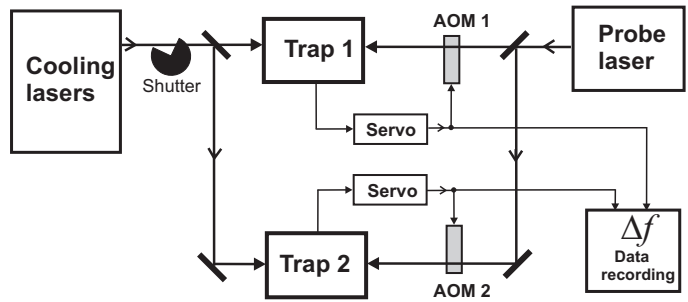}
\end{center}
\caption{Experimental setup for the comparison of two $^{171}$Yb$^+$ frequency standards. AOM: acusto-optic modulators, providing independent frequency shifts between the probe laser and the two ion traps.}
\end{figure}

\section{Frequency comparisons between two trapped ions}

To enable a quantitative study of systematic frequency shifts on the sub-hertz scale,
we compare the line center frequencies of two $^{171}$Yb$^+$ ions stored in separate traps.
The scheme of the frequency comparison
experiment is shown in Fig. 2. Both traps use the same cooling laser setup and synchronous timing schemes
for cooling, state preparation, and state detection.
The beam from the probe laser is split and  two
separate frequency shift and servo systems are employed to stabilize the probe frequencies to the reference transition line centers of the two ions.
The atomic resonance signal is probed in typically eight excitation cycles until an error signal is calculated and the detunings between the probe laser frequency and the probe light beams incident on the traps are corrected.
The servo time constants are in the range of 10 s.
In order to minimize servo errors due to the drift of the probe laser frequency, a second-order integrating servo algorithm is used.  The differences of the detunings imparted on the probe beams are averaged over time intervals of 1 s and recorded.
As a measure of the instability, the Allan deviation of a typical data set is shown in Fig. 3.
The result indicates a relative instability of the frequency difference between the two systems of $\sigma_y(1000\,{\rm s})=5\cdot 10^{-16}$.
The atomic resonance signals were resolved with Fourier-limited linewidths of approximately 30 Hz in both traps.
For averaging times that are longer than the time constants of the servo systems, the so-called quantum projection
noise \cite{qpn} limits the instability. This contribution to the instability is dominant because 
the state measurement is done on a single atom only.
We have performed a numerical calculation which simulates the effect of quantum projection noise  for the realized experimental conditions and include the result as the solid line in Fig. 3, showing good agreement with the measured data.

\begin{figure}[tbh]
\begin{center}
\includegraphics[width=8cm]{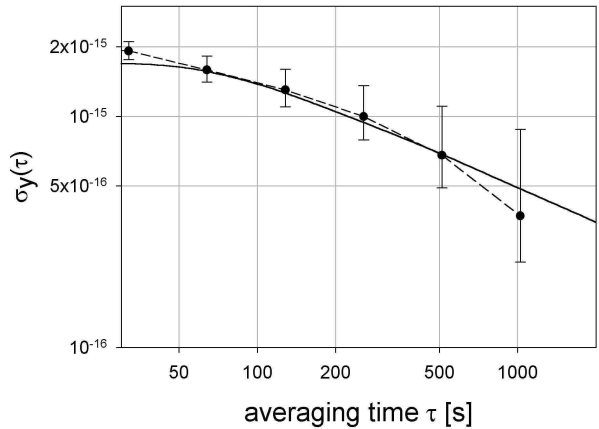}
\end{center}
\caption{Allan standard deviation of the frequency differences observed between the two traps, normalized to the optical frequency of 688 THz. The dashed line connects the experimental points and the solid line shows the result of a Monte Carlo simulation of the servo action for the case that the fluctuations of the atomic resonance signals are determined by quantum projection noise.}
\end{figure}

The mean frequency difference between the two traps was generally in the hertz or sub-hertz range for nominally unperturbed conditions \cite{icols}.
The dominant source of systematic uncertainty in the $^{171}$Yb$^+$ optical frequency standard is the so-called 
quadrupole shift of the atomic transition frequency. It is due to the interaction
of the electric quadrupole moment of the $D_{3/2}$ state with the gradient of the static electric field 
in the trap.
In order to experimentally determine the $^{171}$Yb$^+$ quadrupole moment, a static field gradient was generated in one of the traps by superimposing a constant voltage on the radiofrequency trap drive voltage. The orientation of this field gradient is determined by the symmetry axis of the trap. The other trap was operated with a pure rf voltage and served as a reference. From this experiment, we infer a quadrupole moment of $Q = (3.9 \pm  1.9)ea_0^2$ for the $^2D_{3/2}$ level with $e$ being the elementary charge and $a_0$ the Bohr radius. The uncertainty is mainly determined by the uncertainty of the measurement of the angle between the magnetic field and the trap axis.
Depending on the orientation, a field gradient of 1~V/mm$^2$ may produce a frequency shift of up to 6~Hz.
Averaging the frequency over three orthogonal quantization axes -- i.e. axes of the magnetic field --
allows to determine the unperturbed transition frequency \cite{itano}.

\begin{figure}[tbh]
\begin{center}
\includegraphics[width=9cm]{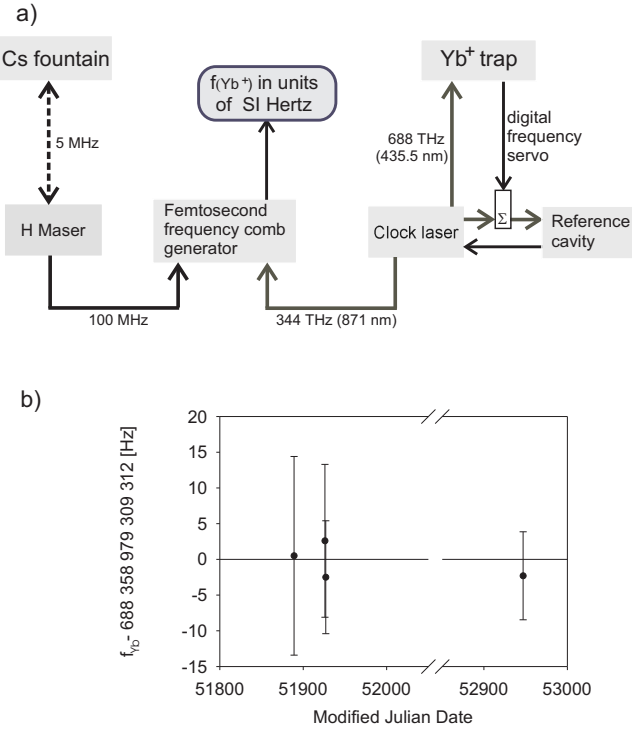}
\end{center}
\caption{a) Schematic of the setup for the absolute frequency measurements of the $^{171}$Yb$^+$ frequency standard.
b) Results of absolute frequency measurements of the $^{171}$Yb$^+$ standard versus the date of the measurement
(MJD: Modified Julian Date).}
\end{figure}

\section{Absolute frequency measurements}

The absolute frequency of the $^{171}$Yb$^+$ standard at 688 THz was  measured relative to a cesium fountain clock \cite{BauchCs},
using the setup shown schematically in Fig. 4a. A hydrogen maser is referenced to the cesium clock and delivers a signal at 100 MHz with a relative instability below $2\cdot 10^{-15}$ in 1000 s of averaging.
The link between
the optical and microwave frequencies is established 
with a femtosecond-laser frequency comb generator. 
The $^{171}$Yb$^+$ frequency was measured 
first in December 2000 and January 2001.
The result was $688\, 358\, 979\, 309\, 312$~Hz with a total $1\sigma$ measurement uncertainty of 
$6$ Hz \cite{Stenger1}. In November 2003,
the frequency was remeasured after improvements in the resolution of the ionic resonance and 
in the frequency comb generator setup. The new value has a total $1\sigma $ uncertainty of 6.2~Hz and is insignificantly lower by 2.3~Hz. 
The dominant source of systematic uncertainty is the quadrupole shift
that may be produced by patch charges on the trap electrodes.
The static quadrupole potential associated with these charges
will also alter the oscillation frequencies of the ion in the trap.
Measurements of the oscillation frequencies allow
to estimate the uncertainty due to the quadrupole shift as 3 Hz under the conditions of the experiment. 
The total uncertainty of 6.2~Hz (corresponding to a relative uncertainty of $9\cdot 10^{-15}$) contains also
a systematic contribution of 2.2~Hz from the microwave reference and frequency comb generator and a total statistical contribution of 4.8~Hz.

From the sequence of absolute frequency measurements (cf. Fig. 4b) we deduce
a value for the fractional temporal variation of the frequency ratio $f_{\rm Yb}/f_{\rm Cs}$ of
$(-1.2\pm 4.4)\cdot 10^{-15}$~yr$^{-1}$, consistent with zero. This result can be used to obtain a limit on the temporal variation of the fine structure constant.

\section{Variations of fundamental constants}

Discussions about a  possible temporal variability of fundamental constants of nature
date back until the 1930s, with Dirac's large number hypothesis \cite{dirac} being a prominent example. Meanwhile, strong motivation to search for such an effect has developed because theories that attempt to unify the fundamental interactions allow or even imply temporal variations of the coupling constants (see e.g. \cite{barrow}).
Variations of the constants have been searched for in various systems \cite{acfc,uzan}.
Sommerfeld's fine-structure constant $\alpha=e^2/4\pi \epsilon_0 \hbar c$ (composed of the elementary charge $e$, Planck's constant $\hbar$, the speed of light $c$, and the electric constant of vacuum $\epsilon_0$) is presently the most important test case because it is the fundamental coupling constant of the electromagnetic interaction. As a dimensionless quantity $\alpha \approx 1/137.036$ can be determined without reference to a specific system of units.
This point is important because the realization of the units themselves may also be affected by variations of the constants.

An indication for a possible variation of $\alpha$ comes from a shift of wavelengths of specific absorption lines produced by interstellar clouds in the light from distant quasars \cite{webb1}. These observations
seem to suggest that about $1\cdot 10^{10}$~yr ago
the value of $\alpha$ was smaller than today by $\Delta \alpha / \alpha= (-0.543\pm 0.116)\cdot 10^{-5}$, 
representing  $4.7\sigma$ evidence for a varying $\alpha$ \cite{webb2}.
Assuming a linear increase in $\alpha$ with time, this would correspond to a drift rate
$\partial \ln \alpha / \partial t=(6.40\pm 1.35)\cdot 10^{-16}~{\rm yr}^{-1}$ \cite{webb2}.
Recent evaluations of new data on quasar absorption lines are consistent with $\Delta \alpha=0$  for all look-back times \cite{sria}. 

To perform a laboratory tests for the variability of $\alpha$ in the present  epoch
one can compare different electronic transition frequencies in the optical range \cite{Karsh,Karsh2}.
For the analysis, we express the electronic transition frequency as 
\begin{equation}
f=Ry\cdot C\cdot F(\alpha)
\end{equation}
where 
$Ry=m_e e^4/(8\epsilon_0 h^3)\simeq 3.2898\cdot 10^{15}$ Hz 
is the Rydberg constant in SI units of frequency, appearing here as the common atomic scaling factor.
$C$ is a numerical constant which depends only on the quantum numbers characterising the atomic state
and which is independent of time. $F(\alpha)$ is a dimensionless function of $\alpha$
that takes into account level shifts due to relativistic effects \cite{dzu}. 
In  many-electron atoms these effects lead to different sensitivities of the transition frequencies to a change of $\alpha$, with the general trend to find a stronger dependence in heavier atoms \cite{prest}. 
The relative temporal derivative of the frequency $f$ can be written as:
\begin{equation}
\frac{\partial \ln f}{\partial t} =  \frac{\partial \ln Ry}{\partial t} + A \cdot \frac{\partial \ln \alpha}{\partial t}~~~~~{\rm with~}A\equiv \frac{\partial \ln F}{\partial \ln \alpha}.
\end{equation}       
The first term $\partial \ln Ry / \partial t$ represents a variation that would be common to all measurements of electronic transition frequencies: a change of the numerical value of the Rydberg constant in SI units, i.e.
with respect to the $^{133}$Cs hyperfine splitting.
We distinguish here between the numerical value of $Ry$ and the physical parameter $m_e e^4/h^2$ which determines the atomic energy scale. A change of the latter would not be detectable in this type of measurement because the   
$^{133}$Cs hyperfine splitting also scales with this quantity.
Consequently, the detection of a non-zero value of $\partial \ln Ry / \partial t$ would have to be interpreted as being related to a change of the nuclear magnetic moment of $^{133}$Cs.
The second term in Eq. 2 is specific to the atomic transition under study.
The  sensitivity factor $A$ for small changes of $\alpha$ has been calculated for a number of cases
of interest by Dzuba, Flambaum et al. \cite{dzu}.         
Assuming that all possible drifts are linear in time over the duration of the frequency
comparisons, the fractional drift rate $\partial \ln \alpha / \partial t$ can be obtained from at least two measured drift rates of transition frequencies $\partial \ln f / \partial t$ if these have been measured against a common reference, and if the sensitivity factors $A$ for the two transitions are different. 

To derive the limit on $\partial \ln \alpha / \partial t$ we combine the data from $^{171}$Yb$^+$ with a published 
limit on the drift rate of a transition frequency in $^{199}$Hg$^+$ 
\cite{Bize}, obtained at NIST (Boulder, USA).
In Hg$^+$ the transition $5d^{10}6s~^2S_{1/2}\rightarrow 5d^96s^2~^2D_{5/2}$ at 282~nm (1065 THz) with a natural linewidth of 1.9~Hz serves as the reference. 
A first absolute frequency measurement of this transition with respect to a cesium clock was published in 2001 with an uncertainty of $9\cdot 10^{-15}$ \cite{Udem2}. A sequence of measurements over the period from
August 2000 to November 2002
has resulted in a constraint for the fractional variation of  
$f_{\rm Hg}/f_{\rm Cs}$ at the level of $7\cdot 10^{-15}$~yr$^{-1}~(1\sigma)$ \cite{Bize}. 
The sensitivities of the ytterbium and mercury transition frequencies to changes of  $\alpha$ are quite different: $A_{\rm Yb}=0.88$ and $A_{\rm Hg}=-3.19$ \cite{dzu}. 

A further precision test for the constancy of an optical transition frequency was presented by
Fischer et al. who measured the $1S\rightarrow 2S$ transition frequency in atomic hydrogen in two periods in July 1999 and in February 2003, using a transportable cesium fountain frequency standard  \cite{fischer}.
The limit on the fractional drift rate for the hydrogen frequency that is deduced from these experiments is $(-3.2\pm 6.3)\cdot 10^{-15}$~yr$^{-1}$ \cite{fischer} with the sensitivity factor $A_{\rm H}$ being zero.

\begin{figure}[tbh]
\begin{center}
\includegraphics[width=6cm]{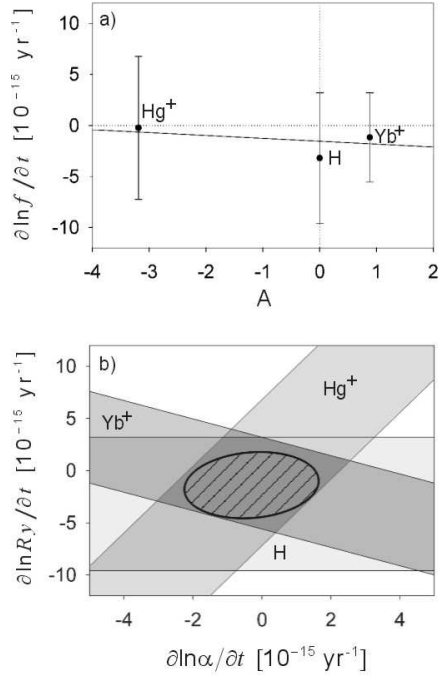}
\end{center}
\caption{a) Relative frequency drift rate versus sensitivity factor $A$ for the $S\rightarrow D$ reference transitions in the Hg$^+$ and Yb$^+$ ions and for the $1S\rightarrow 2S$ transition in atomic hydrogen. 
The straight line is the result of a weighted linear regression.
A significant deviation of the slope from zero would indicate a non-zero 
time derivative of the fine structure constant $\alpha$.
b) Constraints on the variability of $\alpha$ and the numerical value of $Ry$ as obtained from the experimental points in (a). 
Stripes mark the 1$\sigma$ uncertainty of the individual measurements
and the central hatched area is bounded by the standard uncertainty ellipse resulting from the 
combination of all three experiments.}
\end{figure}

Plotting the three limits on the frequency drift rates as a function of $A$ (Fig. 5a) and performing a 
weighted linear regression 
we obtain from the slope of the fit:
\begin{equation}
\frac{\partial \ln \alpha}{\partial t} = (- 0.3 \pm 2.0)\cdot 10^{-15}~{\rm yr}^{-1}
\end{equation}
for the present value of the temporal derivative of the fine structure constant at the confidence level of $1\sigma$. 
This limit is 
based on a model-independent analysis, not using assumptions on correlations between drift rates of different quantities and not postulating the drift rate of any quantity to be zero. To our knowledge, this is so far the most stringent limit on $\partial \alpha/ \partial t$ obtained from atomic clock comparisons \cite{peik}.
Another related analysis based on the data from hydrogen and mercury is presented in \cite{fischer}.

The intercept of the straight line representing the dependence of the optical frequency drift rate on $A$ with the line $A=0$ in Fig. 5a  can be used to determine the drift rate of the numerical value of the Rydberg 
frequency.
This quantity is of great metrological importance.
A non-zero value would imply among other consequences that a time scale generated with a cesium clock
would not be uniform with respect to other atomic oscillations.
The result that we obtain here is consistent with zero:
\begin{equation}
\frac{\partial \ln Ry}{\partial t}=(-1.6 \pm 3.2)\cdot 10^{-15}~{\rm~yr}^{-1}.
\end{equation}
Fig. 5b shows how the measurements of the three transition frequencies contribute to the constraints on the temporal derivatives of $\alpha$ and $Ry$. The central hatched area is bounded by the standard uncertainty ellipse (see Ref. \cite{peik} for details).

With these results we reach a sensitivity for $\Delta \alpha /\Delta t$ with $\Delta t\approx 3$~yr that approaches that of the quasar observations \cite{webb2,sria} to within about a factor of three if a linear change of $\alpha$ is assumed on a cosmological time scale.
Further improvements in the data from clock comparisons can be expected since trapped ions have the potential
to reach 
relative uncertainties in the range of $10^{-18}$ for the comparison of optical transition frequencies \cite{Madej}.
The kind of analysis that we have applied here to the frequencies of the $S\rightarrow D$ quadrupole transitions in $^{171}$Yb$^+$ and $^{199}$Hg$^+$ and of the $1S\rightarrow 2S$ transition in H may be extended to other atomic systems as soon as precision data on frequency drift rates become available.

\bigskip
\noindent
{\bf Acknowledgements}\\

We are grateful to A.~Bauch, S. Weyers and R. Wynands for contributions to the experiments and helpful discussions. This work was partially supported by DFG through SFB 407 and by RFBR under grants 03-02-04029.


\begin{thebibliography}{99}

\bibitem{Madej}
A. A.~Madej, J. E.~Bernard, 
in: {\it Frequency Measurement and Control:
Advanced Techniques and Future Trends},  Topics in Applied
Research, Ed.: A. N.~Luiten (Springer, Berlin, Heidelberg, 2000).

\bibitem{Gill} P. Gill et al., Meas. Sci. Tech. {\bf 14}, 1174 (2003).

\bibitem{Reichert}
J.~Reichert, M.~Niering, R.~Holzwarth, M.~Weitz, Th.~Udem,
T.W.~H\"ansch, Phys. Rev. Lett. {\bf 84}, 3232 (2000);
S.A.~Diddams, D.J.~Jones, J.Ye, S.T.~Cundiff, J.L.~Hall,
J.K.~Ranka, R.S.~Windeler, R.~Holzwarth, Th.~Udem, T.W.~H\"ansch,
Phys. Rev. Lett. {\bf 84}, 5102 (2000).


\bibitem{Bernard}
J. E.~Bernard, A. A. Madej, L.~Marmet, B. G. Whitford, K. J. Siemsen, S. Cundy, Phys. Rev. Lett. {\bf 82}, 3228
(1999).

\bibitem{Margolis} H. S. Margolis et al., Phys. Rev. A {\bf 67}, 032501 (2003).

\bibitem{Zanthier}
J.~v.~Zanthier et al.,  Opt. Lett. {\bf 25}, 1729 (2000).

\bibitem{Becker}
Th.~Becker, J.~von~Zanthier, A. Yu.~Nevsky, Ch.~Schwedes,
M. N.~Skvortsov, H.~Walther, E.~Peik, Phys. Rev. A  {\bf 63},
051802(R) (2001).

\bibitem{Stenger1} J. Stenger, Chr. Tamm, N. Haverkamp, S. Weyers, and H. Telle, Opt. Lett. {\bf 26} 1589 (2001).

\bibitem{Tamm1}
Chr. Tamm, D. Engelke, and V. B\"uhner, Phys. Rev. A {\bf 61}, 053405 (2000).

\bibitem{Tamm2} Chr. Tamm, T. Schneider, and E. Peik, in: {\it Proc. of the 6. Symp. on Frequency Standards and Metrology}, Ed.: P. Gill (World Scientific, Singapore,  2002).

\bibitem{Blythe} P. Blythe et al., Phys. Rev. A {\bf 67}, 020501(R) (2003).

\bibitem{Rafac}
R. J. Rafac, B. C. Young, J. A. Beall, W.~M. Itano, D.~J. Wineland,
and J.~C. Bergquist, Phys. Rev. Lett. {\bf 85},  2462 (2000).

\bibitem{Bize} S. Bize et al., Phys. Rev. Lett. {\bf 90}, 150802 (2003).

\bibitem{icols} Chr. Tamm, T. Schneider, and E. Peik, in: 
Laser Spectroscopy XVI, Eds.: P. Hannaford, A.
Sidorov, H. Bachor, and K. Baldwin (World Scientific, Singapore, 2004). 

\bibitem{peik}
E. Peik, B. Lipphardt, H. Schnatz, T. Schneider, Chr. Tamm,
S. G. Karshenboim, Phys. Rev. Lett. {\bf 93}, 170801 (2004).

\bibitem{qpn} W. M. Itano et al., Phys. Rev. A {\bf 47}, 3554 (1993).

\bibitem{itano} W. M. Itano, J. Res. NIST {\bf 105}, 829 (2000).

\bibitem{BauchCs} A. Bauch, Meas. Sci. Technol. {\bf 14}, 1159 (2003).

\bibitem{dirac} P. A. M. Dirac, Nature {\bf 139}, 323 (1937).

\bibitem{barrow} J. D. Barrow, Astrophys. Space Sci. {\bf 283}, 645 (2003).

\bibitem{acfc} {\it Astrophysics, Clocks and Fundamental Constants}, Lecture Notes in Physics Vol. 648,  Eds.: S. G. Karshenboim and E. Peik
(Springer, Heidelberg, 2004).

\bibitem{uzan} J.-P. Uzan, Rev. Mod. Phys. {\bf 75}, 403 (2003). 

\bibitem{webb1} J. K. Webb et al., Phys. Rev. Lett. {\bf 87}, 091301 (2001).

\bibitem{webb2} M. T. Murphy, J. K. Webb, V. V. Flambaum, Mon. Not. R. Astron. Soc. {\bf 345}, 609 (2003).

\bibitem{sria}R. Quast, D. Reimers, S. A. Levshakov, Astr. Astrophys. {\bf 415}, L7 (2004); R. Srianand, H. Chand, P. Petitjean, B. Aracil, Phys. Rev. Lett. {\bf 92}, 121302 (2004).

\bibitem{Karsh} S. Karshenboim, Can. J. Phys. {\bf 78}, 639 (2001).

\bibitem{Karsh2} S. G. Karshenboim, eprints physics/0306180 and physics/0311080.

\bibitem{dzu} V. A. Dzuba, V. V. Flambaum, J. K. Webb, Phys. Rev. A {\bf 59}, 230 (1999);
V. A. Dzuba, V. V. Flambaum, Phys. Rev. A {\bf 61}, 034502 (2000);
V. A. Dzuba, V. V. Flambaum, M. V. Marchenko, Phys. Rev. A {\bf 68}, 022506 (2003).

\bibitem{prest} J. D. Prestage, R. L. Tjoelker, L. Maleki, Phys. Rev. Lett. {\bf 74}, 3511 (1995).

\bibitem{Udem2} Th. Udem et al.,  Phys. Rev. Lett. {\bf 86}, 4996 (2001).

\bibitem{fischer} M. Fischer et al., Phys. Rev. Lett. {\bf 92}, 230802 (2004).

\end{thebibliography}
\end{document}